\documentclass[%
 reprint,
 amsmath,amssymb,
 aps,
]{revtex4-1}

\usepackage{graphicx,txfonts}
\usepackage{chemarr}
\usepackage{times,longtable}
\usepackage{hyperref}
\usepackage{version}
\usepackage{color}
\usepackage{soul}

\usepackage{epsf,mathtools}
\usepackage[usenames,dvipsnames]{xcolor}

\renewcommand{\mod}[1]{\textcolor{black}{#1}}

\newcommand{\modminor}[1]{\textcolor{black}{#1}}

\begin{document}

\title[Coevolutionary mutational landscape inference]{
Coevolutionary landscape inference\\ and the context-dependence of mutations in beta-lactamase TEM-1}

\author{M. \surname{Figliuzzi},$^{1,2,3}$ H. Jacquier,$^{4,5}$ A. Schug,$^{6}$ O. Tenaillon,$^{4}$ M. Weigt$^{\ast,2,3}$\\
\vspace*{0.5cm}
{\it $^{1}$Sorbonne Universit\'es, UPMC, Institut de Calcul et de la Simulation, Paris, France\\
$^{2}$Sorbonne Universit\'es, UPMC, UMR 7238, Computational and Quantitative Biology, Paris, France\\
$^{3}$CNRS, UMR 7238, Computational and Quantitative Biology, Paris, France\\
$^{4}$INSERM, IAME, UMR 1137, Paris, France\\
Universit\'e Paris Diderot, IAME, UMR 1137, Sorbonne Paris Cit\' e, Paris, France\\
$^{5}$Service de Bact\'eriologie-Virologie, Groupe Hospitalier Lariboisi\'ere-Fernand Widal, Assistance Publique-H\^{o}pitaux de Paris (AP-HP), Paris, France\\
$^{6}$Steinbuch Centre for Computing, Karlsruhe Institute for Technologie, Eggenstein-Leopoldshafen, Germany\\
$^{\ast}$ E-mail: martin.weigt@upmc.fr}}



\begin{abstract}
The quantitative characterization of mutational landscapes is a task of outstanding importance in evolutionary and medical biology: It is, e.g., of central importance for our understanding of the phenotypic effect of mutations related to disease and antibiotic drug resistance. Here we develop a novel inference scheme for mutational landscapes, which is based on the statistical analysis of large alignments of homologs of the protein of interest. Our method is able to capture epistatic couplings between residues, and therefore to assess the dependence of mutational effects on the sequence context where they appear. Compared to recent large-scale mutagenesis data of the beta-lactamase TEM-1, a protein providing resistance against beta-lactam antibiotics, our method leads to an increase of about 40\% in explicative power as compared to approaches neglecting epistasis. We find that the informative sequence context extends to residues at native distances of about 20\AA~from the mutated site, reaching thus far beyond residues in direct physical contact.\\
\vspace*{0.2cm}

\end{abstract}

\maketitle

\section*{Introduction\label{sec:Intro}}

Protein mutational landscapes are genotype-to-phenotype mappings quantifying how mutations affect the biological functionality of a protein. They are closely related to fitness landscapes describing the replicative capacity of an organism as a function of its genotype \cite{wright1932roles}. Their comprehensive and accurate characterization is a task of outstanding importance in evolutionary and medical biology: It has a key role in our understanding of mutational pathways accessible in the course of evolution \cite{kauffman1987towards,weinreich2006darwinian,poelwijk2007empirical}, it can lead to the identification of genetic determinants of complex diseases based on rare variants \cite{cirulli2010uncovering}, and it can guide towards the understanding of the functional contribution of molecular alterations to oncogenesis \cite{reva2011predicting}. 
In the context of antibiotic resistance, one of the most challenging problems in modern medicine, the understanding of the association between genetic variation and phenotypic effects can help to unveil patterns of adaptive mutations of the pathogens to gain drug resistance, and thereby hopefully guide toward the discovery of new therapeutic strategies \cite{ferguson2013translating}.

One key issue in the description of a mutational landscape is to understand how much the effect of a mutation depends on the genetic background in which it appears \cite{weinreich2006darwinian,khan2011negative,chou2011diminishing}. For instance, in the field of human genetic diseases, is the presence of a mutation enough to predict a pathology or do we have to know the whole genotype to make that assertion? In a more formal way, this question is equivalent to quantifying how epistasis, i.e. the interaction between mutations through fitness, is shaping the mutational landscape. At the protein level, a destabilizing mutation might have a negligible phenotypic effect in a very stable protein, but a large one in an unstable protein \cite{bloom2005thermodynamic,jacquier2013capturing}. If \mod{this destabilizing} mutation increases, e.g., the enzymatic activity, it will be beneficial in a stable protein, and deleterious in an unstable one, cf.~\cite{harms2013evolutionary}.
Hence the mutation is expected to be context dependent. Moreover, once a mutation has fixed, further mutations will build upon the specificity of that focal mutation, thereby creating a new genetic background with its specific interactions and interdependencies \cite{pollock2012amino}. There are ample proofs of the existence of epistasis and condition dependent effects \cite{harms2013evolutionary,breen2012epistasis,de2014empirical,Podgornaia06022015,schenk2013patterns}. Yet, it is not totally clear whether such interactions have a dominant or a minor effect in determining a mutation's phenotypic impact.

Recent technological advances have made it possible to simultaneously quantify the effects of thousands to hundreds of thousands of mutants through either growth competition \mod{\cite{mclaughlin2012spatial,deng2012deep,Podgornaia06022015,melamed2013deep,roscoe2013analyses}} or isolated allele experiments \mod{\cite{jacquier2013capturing,firnberg2014comprehensive,romero2015dissecting}}. Experimental resolution can be good enough to detect even the effects of synonymous mutations \cite{firnberg2014comprehensive}. Despite the development of such high-throughput methods, measured genotypes cover only a tiny fraction of sequence space: The number of possible mutants grows exponentially with the number of single mutations, such that checking the viability of all possible genotypes further than one or two mutations away from a reference sequence becomes infeasible, even for short polypeptides. More precisely, the number of distinct single-residue mutants for typical proteins is in the range of $10^3-10^4$. The number of all double mutants reaches the range of $10^6-10^8$. While this number is not yet experimentally accessible, it is needed to accurately assess the importance of epistasis. It has been argued that existing mutagenesis data are not sufficient for accurate landscape regression \cite{Otwinowski03062014}. Novel computational approaches exploring alternative data -- in our case distant homologs -- are thus urgently needed to gain a comprehensive picture of mutational landscapes. In this context, the growing amount of mutagenesis data offers the possibility to rigorously evaluate the performance of {\em in-silico} models of mutational landscapes.

Several computational methods for predicting mutational effects on protein function have been proposed over the years. A first class relies on {\em structural} information, more precisely on changes in the thermodynamic stability \cite{cheng2006prediction,capriotti2005mutant2,lonquety2009sprouts,dehouck2011popmusic,
ng2006predicting,capra2007predicting}, which have been argued to play a key role in determining mutational effects \mod{\cite{wylie2011biophysical,serohijos2014merging,bloom2009inferring,echave2015relationship}}. A second class \cite{adzhubei2010method,ng2003sift} relies on {\em evolutionary} information extracted from independently evolving homologous proteins, showing variable amino-acid sequences but conserved structure and function. Evolution provides a multitude of informative 'experiments' on mutational landscapes. Critically important residues tend to be conserved, while unfavorable residues are observed less frequently. 

None of these methods is able to model the effects of epistasis and sequence-context dependence of mutational effects. 
To overcome this limitation, we take inspiration from a recent development in structural biology. It has been recognized that coevolutionary information contained in large families of homologous proteins allows to extract accurate structural information from sequences alone \cite{de2013emerging}: Residues in contact in a protein's fold, even if distant along the primary sequence, tend to show correlated patterns of amino-acid occurrences. Inversely, correlated residues are not necessarily in contact, since correlations are inflated by indirect effects. Two residues, both being in contact to a third residue, will coevolve even if they are not in direct contact. The Direct-Coupling Analysis (DCA) \cite{weigt2009identification,morcos2011direct} has been proposed to disentangle such indirect effects from direct (i.e. epistatic) couplings, which in turn have been observed to accurately predict residue-residue contacts. DCA and closely related methods thereby guide tertiary \cite{marks2012NATBIOTECH,sulkowska2012genomics,hopf2012three,nugent2012accurate} and quaternary \cite{schug2009high,dago2012structural,ovchinnikov2014robust,hopf2014elife} protein structure prediction; and shed light on specificity and crosstalk in bacterial signal transduction \mod{\cite{procaccini2011dissecting,cheng2014toward}}.

In this paper we propose a variant of DCA which assigns to each mutant sequence a statistical score, which in a next step is used for predicting the phenotype of the mutant sequence relative to the wild-type sequence. To evaluate the approach, we take the {\it Escherichia coli} beta-lactamase TEM-1, a model enzyme in biochemistry which provides resistance to beta-lactam antibiotics. Its mutational landscape has been quantitatively characterized measuring the minimum inhibitory concentration (MIC) of the antibiotic \cite{davison2000antibiotic,jacquier2013capturing,firnberg2014comprehensive}. This abundance of mutagenesis data, the rich homology information and its well defined 3D structure make it a well-suited system for testing any computational model of protein mutational landscapes. 

We will show that coevolutionary models for mutational landscapes do not only provide quantitative predictions of mutational effects but, more importantly, they are able to capture the context dependence of these effects. In this way, the new approach manages to clearly outperform state-of-the-art approaches like SIFT \cite{ng2003sift} and PolyPhen-2 \cite{adzhubei2010method}, which are based on independent-site models (even if, like in the case of PolyPhen-2, additional structural information is integrated into the prediction of mutational effects), which themselves outperform predictors based on structural stability.
The approach is broadly applicable, \mod{as is illustrated in a small set of completely different systems: a RNA recognition motif \cite{melamed2013deep}, the glucosidase enzyme \cite{romero2015dissecting} and a PDZ domain \cite{mclaughlin2012spatial}. In the last system positions most sensitive to mutation had been shown previously to fall into clusters of coevolving residues termed sectors \cite{halabi2009protein}}: Appling statistical inference we are able to get a more {\em quantitative} prediction of the impact of single point mutations in the domain.
These findings illustrate the potential of coevolutionary landscape models in biomedical applications, via the {\em in-silico} prediction of mutational effects not only related to antibiotic drug resistance, but also to the role of mutations in rare diseases and cancer.

\section*{Results\label{sec:Results}}

\subsection*{Evolutionary modeling of diverged beta-lactamase sequences to predict 
mutational effects of single-residue mutations in TEM-1}

\begin{figure}
\begin{center}
\includegraphics[width=.9\textwidth]{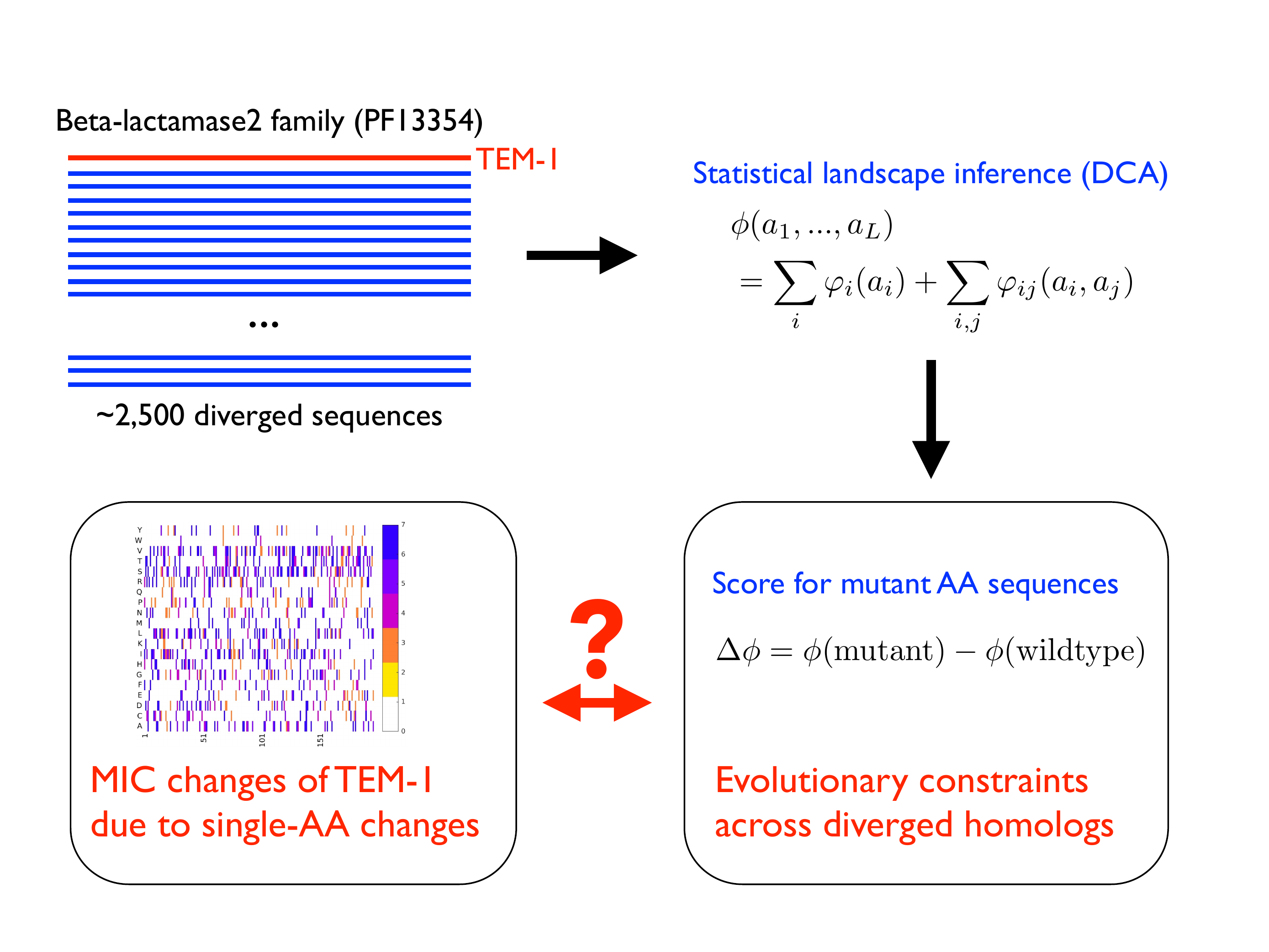}
\vspace*{0.05cm}
\caption{\label{fig:pipeline}\mod{ {\bf Pipeline of the mutational-landscape prediction}: 
The homologous Pfam family containing the protein of interest (the Beta-lactamase2 
family PF13354 in the case of TEM-1) is used to construct a global statistical model
using the Direct-Coupling Analysis (DCA). This model allows to score mutations by
differences in the inferred genotype-to-phenotype mapping between the mutant and 
the wild-type amino-acid sequence. This score, which is expected to incorporate 
(co-)evolutionary constraints acting across the entire family, is used as a predictor 
of the phenotypic effects of single (or few) amino-acid substitutions in the protein of interest.} }
\end{center}
\end{figure}

The pipeline of our approach is illustrated in Fig. \ref{fig:pipeline}.

\onecolumngrid

\twocolumngrid

In technical terms, a mutational landscape is given as a genotype-to-phenotype mapping. To each possible amino-acid sequence $(a_1,...,a_L)$ consisting of $L$ amino acids or gaps ($L$ denotes the alignment width), a quantitative phenotype $\phi(a_1,...,a_L)$ is assigned. The phenotypic effect of a mutation substituting the wild-type amino acid $a_i$ at position $i$ with amino acid $b$ is measured by the difference score
\begin{align}
& \Delta \phi (a_i\to b) = 
\phi(a_1,...,a_{i-1},b,a_{i+1},...,a_L) - \nonumber \\ 
 & - \phi(a_1,...,a_{i-1},a_i,a_{i+1},...,a_L)\
\label{eq:Dphi}
\end{align}
between the mutant and the wild-type sequence. This function $\phi$ has, however, $20^L$ parameters, an astronomic number being far beyond any possibility of inference from data. Simplified parameterizations of $\phi$ reducing the number of parameters are needed. In general, a simple model can be inferred more robustly from limited data, but it risks to miss important effects. Even if these might be captured in more complex models, the latter risk to suffer from undersampling and thus overfitting effects. One of our aims is to find a good compromise between these two limitations.

The simplest non-trivial parametrization assumes position-specific but independent contributions of each residue, 
\begin{equation}
\phi_{IND}(a_1,...,a_L) = \sum_{i=1}^L \varphi_i(a_i) \  .
\label{eq:IND}
\end{equation}
The contribution $\varphi_i(a_i)$ measuring the contribution of amino acid $a_i$ in position $i$ can be easily estimated from a multiple-sequence alignment (MSA) of homologous proteins using the framework of profile models (also called position-specific weight matrices), cf. {\em Methods} for details. Possibly existing epistatic effects are neglected. Within this modeling scheme, the score for a single amino-acid substitution simplifies from Eq. (\ref{eq:Dphi}) to $\Delta \phi_{IND} (a_i\to b) = \varphi_i(b)-\varphi_i(a_i)$. It becomes immediately evident that the independent-residue model is unable to capture the context dependence of mutations, the substitution  $a_i\to b$ is predicted to have identical effects if introduced into different sequence backgrounds. The score of a double mutation is simply given by the sum of the $\Delta\phi$-values of the two single-residue mutations.

The relation between statistically derived scores $\Delta\phi$ and the experimental MIC values may be nonlinear. The discrete nature of the latter introduces saturation effects, in particular for strongly deleterious mutations with MIC values below the lowest measured antibiotic concentration. To address these issues, we have designed a robust mapping of $\Delta\phi_{IND}(a_i\to b)$ to predicted MIC values $\hat{\mu}_{IND}(a_i\to b)$, cf. {\em Methods}, and compared them to the experimental MIC values $\mu_{exp}(a_i\to b)$ by linear correlation. A direct measurement of Spearman rank 
correlations between $\phi_{IND}$ and $\mu_{exp}$ leads to numerically very similar, but slightly  less robust results.

The MIC predictions using model Eq. (\ref{eq:IND}) show a Pearson correlation of $R=0.63$ with the experimental
MIC measurements of single-residue substitutions in TEM-1. About $R^2\simeq 39\%$ of the variability of the experimental results is thus explainable by an independent-site model built on the sequence variability between homologous sequences. Very similar correlations ($R^2=0.37$) are found when comparing experimental results and the probabilities of being tolerated as predicted by SIFT, which, like most state-of-the-art methods, is based on conservation profiles in sequence alignments. Higher accuracy is found for PolyPhen-2 ($R^2=0.48$): its improved performance results from the integration of a profile-based score with structural features and amino-acid properties.
 
However, all these predictions are based on the assumption that epistasis between mutations and context dependence can be neglected. The simplest model to challenge this assumption takes into account {\em pairwise epistatic interactions} between different residue positions in the MSA,
\begin{align}
&\phi_{DCA}(a_1,...,a_L) =  \nonumber \\
&= \sum_{i,j=1}^L \varphi_{ij}(a_i,a_j) + \sum_{i=1}^L \varphi_{i}(a_i) \ ,
\label{eq:Pdca}
\end{align}
cf. {\em Methods}.
The terms $\varphi_{ij}(a_i,a_j)$ parametrize the epistatic couplings between amino acids $a_i$ and $a_j$ in aligned positions $i$ and $j$; if they would be set to zero the model would reduce to the independent-site model $\phi_{IND}$. This model has been recently introduced within the {\em Direct-Coupling Analysis} (DCA) of residue coevolution with the aim to infer contacts between residues from sequence information alone, and to enable the prediction of tertiary and quaternary protein structures, cf. the references in the {\em Introduction} of this paper. 

Estimating parameters from aligned sequences is a computationally hard task, but over the last years a number of accurate and computationally efficient approximate algorithms have been developed  \cite{weigt2009identification,morcos2011direct,baldassi2014fast,ekeberg2013improved}. Here we extend the mean-field scheme of Morcos et al. \cite{morcos2011direct}, cf.~{\em Methods}.
\mod{For TEM-1, standard DCA analysis accurately predicts tertiary contacts, cf.~Fig.~S1:
More than 60 non-trivial residue-residue contacts (minimum separation of 5 residues along the sequence) are predicted without error, and more than 200 at a precision of 80\%.}

Having estimated $\phi_{DCA}$ from the MSA, we can follow the same strategy as in the independent-residue case. First, a mutational score is introduced as the difference of the $\phi$-values of the mutated and the wild-type sequences, cf. Eq. (\ref{eq:Dphi}). The inclusion of epistatic couplings leads to an {\em explicit context dependence} of the statistical score of a mutation $a_i \to b$ in position $i$ on all other residues in the wild-type sequences, 
\begin{align}
&  \Delta \phi_{DCA}(a_i\to b\ |\ a_1,...,a_{i-1},a_{i+1},..a_L) = \nonumber \\
& = \varphi_i(b)-\varphi_i(a_i) + \sum_{j=1}^L  \left[ \varphi_{ij}(b,a_j) - \varphi_{ij}(a_i,a_j)
\right]\ .
\label{eq:dcascore}
\end{align}
In a second step, this difference score is mapped to predicted MIC values $\hat{\mu}_{DCA}(a_i\to b)$ and compared to the experimental values $\mu_{exp}(a_i\to b)$ by linear correlation. 

Resulting predictions outperform the independent-residue modeling. DCA-predicted MIC values show a correlation of 
$R=0.74$  with the experimental MIC measurements of single-residue substitutions in TEM-1, i.e.~about $R^2 \simeq 55\%$ of the variability of the experimental results is explained by the DCA-inferred mutational landscape, see Fig.~\ref{fig:corr_bar}, as compared to the 39\% reported before for the IND model. We find that DCA even outperforms the integrative modeling of PolyPhen-2 combining sequence profiles with structural and other prior biological knowledge, demonstrating the power of DCA in capturing epistatic effects in the TEM-1 mutational landscape. 

Applying the same procedure to the data of Firnberg et al. \cite{firnberg2014comprehensive},
which are  highly correlated with the data from Jacquier et al. \cite{jacquier2013capturing} ($R=0.94$), but slightly more precise than that, the correlation is slightly higher ($R=0.76,R^2=0.58$).
Excluding from the analysis those data which display large discrepancies between the two experiments (such discrepancies could be either due to experimental errors or due to antibiotic-specific effects) correlations between our computational score and both datasets rise above $R^2=0.65$, cf. Supplementary Fig.~S2.

We conclude that sequence variability in the Pfam sequence 
alignments of distant homologs is highly informative about the local mutational landscape of TEM-1, despite the low
typical sequence identity of only about 20\%  between the homologs and TEM-1. Moreover, accounting for 
context dependence has a crucial impact on the accuracy of an evolution-based approach, and that global inference methods like DCA can efficiently capture such dependencies.

\subsection*{Assessing the context dependence of mutational effects}

\begin{figure*}
\begin{center}
\includegraphics[width=.85\textwidth]{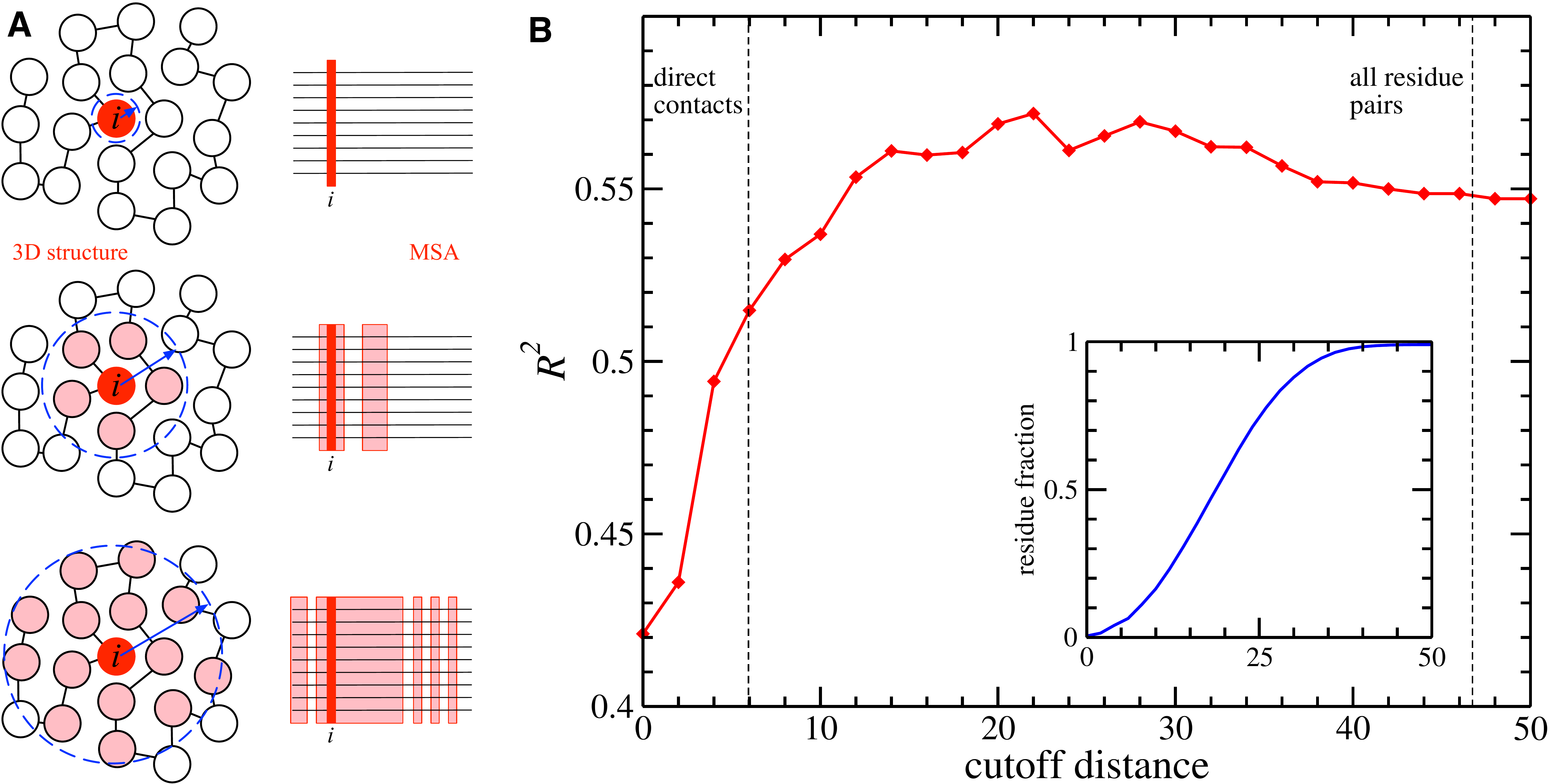}
\vspace*{0.05cm}
\caption{\label{fig:context}   {\bf Context dependence of mutational effects:}
Panel A illustrates the procedure of including all residues within a 
maximal native distance $d_{max}$ into the prediction of the mutational effects of the residue of interest
(labeled $i$ in the figure). This leads to residue-specific sub-alignments, which consist of columns, which are not necessarily consecutive, but connected in 3D. The results are given in Panel B. The main
figure shows the correlation {\scriptsize $R^2$} between MIC data and our predictions, as a function of the
cutoff distance $d_{max}$. The insert shows the average fraction of residues included into the
reduced MSAs, again in dependence of $d_{max}$.}
\end{center}
\end{figure*}

To quantify more precisely the range of context dependence, we apply 
DCA to reduced MSA. These MSA contain the residue position carrying the mutation of interest, and all residues, which 
are, in a representative TEM-1 crystal structure (PDB: 1M40 \cite{minasov2002ultrahigh}), within a distance $d_{max}$ (we use the minimal distance between heavy atoms as the inter-residue distance). When using a very small $d_{max} \leq 1.2$\AA, the mutated residue is considered on its own, when $d_{max}$ is chosen to be larger than the maximum distance 46.9\AA~existing within the PDB structure, we are back to the
full DCA modeling of the previous section. Intermediate $d_{max}$ interpolate between the two 
extreme cases. Doing so, we \mod{run DCA on} sub-alignments of residues, which are not necessarily consecutive in the primary sequence but connected in the native fold, cf.~the illustration 
of the procedure in Panel A of Fig.~\ref{fig:context}. Panel B shows the resulting correlations 
between MIC data and statistical predictions, in function of the cutoff distance $d_{max}$. We 
observe a rapid increase in predictive power when a structural neighborhood is taken into account,
but the increase in correlation extends well beyond the directly contacting residues ($d_{max}\simeq 6$\AA). 
The maximum correlation ($R^2 \simeq 0.57$) is reached around $d_{max} \simeq 20$\AA, followed by a shallow 
decrease when including also more distant residues. This small decrease results probably from
overfitting effects, since the number of model parameters grows quadratically in sequence length.
The insert of Panel B shows the average fraction of residues included into the sub-MSA. At
20\AA~it is slightly higher than 50\%, i.e.~the informative context of a mutation is given by
more than half of the total number of residues in the protein.
\onecolumngrid

\twocolumngrid

It is interesting to observe that \mod{the IND model makes more predictions with very large deviations from the experimental data than the DCA model}: There is an increased number of mutations, which are either predicted to be strongly deleterious even if they are close to neutral, or vice versa. Many of these strong errors are at least partially corrected by the DCA landscape model (cf. Supplementary Tables S1-S3). By the definition of the independent model in terms of frequency counts in individual
MSA sequences, cf. {\em Methods}, a mutation with a low predicted IND score leads from a  
more frequent to a rare amino acid in the concerned MSA column. However, in the
mutagenesis experiments some of these mutations are found to be admissible in
the specific sequence context of TEM-1, i.e. they are actually found to be close to
neutral, examples being G52A, E61V, T112M, N152Y, A183V, T186P, D207V, D250Y 
(all target amino-acids are present in few tens of sequences in the MSA out of the about 
2500 functional homologous sequences). For all of these cases, DCA is able to correct at 
least partially the statistical prediction. 
On the contrary, the independent-site model predicts that any mutation between two
amino acids of similar frequency in the corresponding MSA column is close to 
neutral. Looking to the experimental MIC, substitutions D177N, A235D, I243N and G248E all 
predicted to be close to neutral, have 
strongly deleterious effects (MIC$\leq$25). DCA corrects the mispredictions by at
least two, on average by three MIC classes.

There is a small set of 9 mutations badly predicted by DCA. In none of these cases the independent
modeling significantly ameliorates predictions. Interestingly, 6 out of these 9 mutations fall into the 
highly gapped part of the MSA: DCA display a significant loss of predictive power in the highly 
gapped positions of the MSA, and correlation between predicted and experimental MIC increases 
above $R^2=0.75$ when disregarding mutations in this region (see Supplementary Fig.~S3).

\subsection*{Structural-stability predictions show lower correlations to MIC changes
than sequence-based modeling}

It has been 
proposed before that the role of most residues is to make the protein properly fold, 
and that mutations on these sites mainly alter protein stability and not its activity \cite{wylie2011biophysical}: 
Hence an accurate estimation of the change in protein 
stability $\Delta\Delta G\equiv \Delta G^{mut}-\Delta G^{wt}$ should be able to 
account for a large fraction of mutational effects.

\begin{figure}[h!b]
\begin{center}
\includegraphics[width=.45\textwidth]{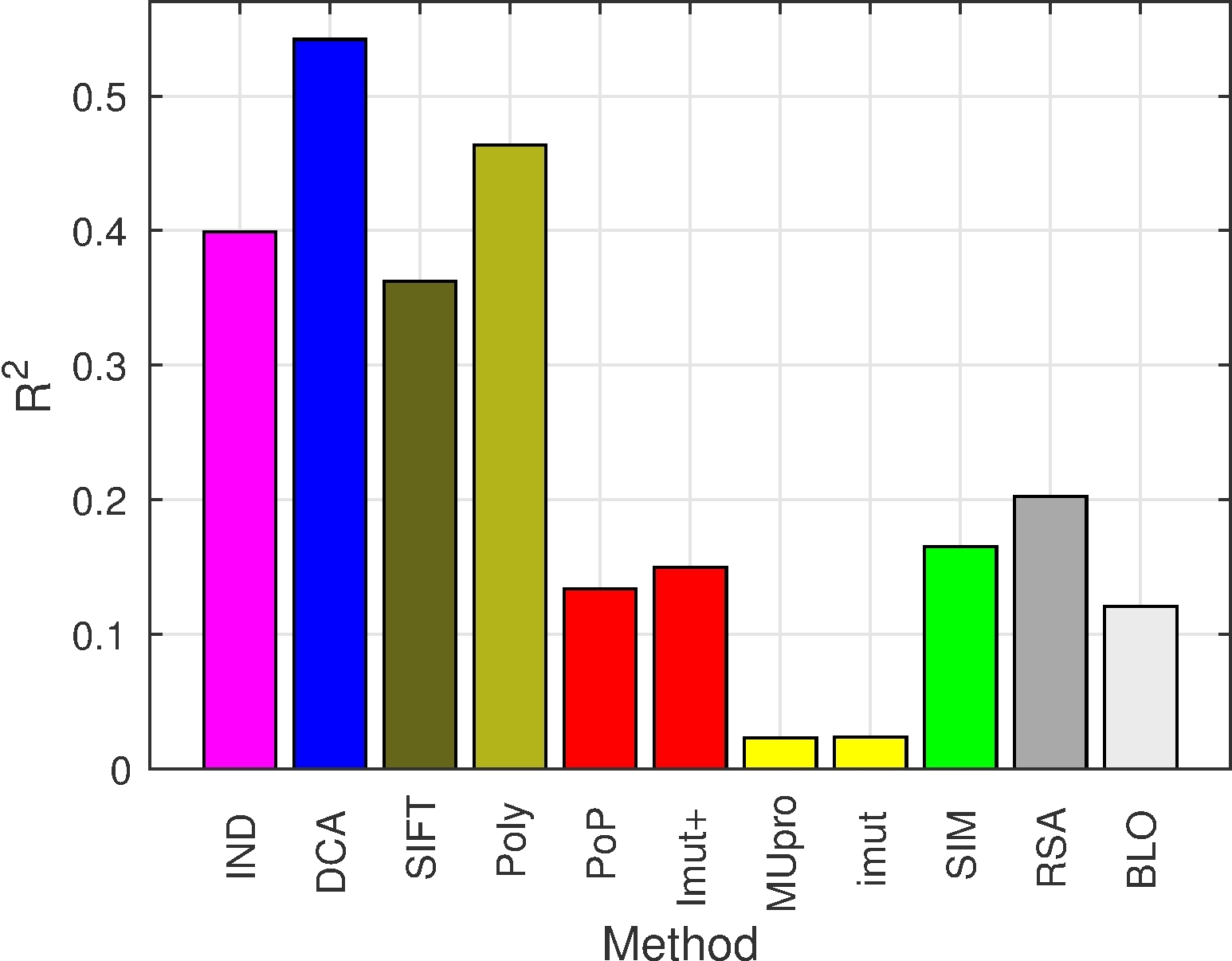}
\vspace*{0.05cm}
\caption{\label{fig:corr_bar} {\bf $R^2$ between experimental fitness and predicted fitness} for the following features:
 Independent-residue model (IND), Direct-Coupling Analysis (DCA), SIFT (SIFT), Polyphen-2 (Poly), PoPMuSiC (PoP), I-Mutant2.0(sequence+structure) (Imut+), MUpro (MUpro), I-Mutant2.0 (Imut), molecular simulations (SIM), \modminor{relative solvent accessibility (RSA)} and Blosum62 substitution matrix (BLO).}
\end{center}
\end{figure}

Many bioinformatic programs have been developed for estimating protein stability 
change upon mutation: among them MUpro \cite{cheng2006prediction} and  
I-Mutant2.0 \cite{capriotti2005mutant2}, which take the sole sequence as input,  
PoPMuSiC \cite{dehouck2011popmusic} and I-Mutant2.0(sequence+structure)\cite{capriotti2005mutant2}, 
which consider both sequence and structure. Since these methods show incoherent 
predictions in between each other, cf.  Supplementary Fig. S4, we complement them 
by extensive force-field molecular simulations at all-atom resolution to estimate protein 
stability changes $\Delta\Delta G$ induced by single point mutations; cf. {\em Methods} for 
details. A score can be assigned 
to any substitution of amino acid $a_i$ in position $i$ by amino acid $b$,
\begin{equation}
\Delta\phi_{stab}(a_i\to b)\equiv -\Delta\Delta G(a_i \to b)\ ,
\end{equation} 
and then mapped to predicted MIC values $\hat\mu_{stab}(a_i\to b)$ using the 
before-mentioned scheme. Pearson correlations between 
predicted and experimental MIC are calculated: We find that, while 
 those methods which consider not only sequence but also structural information 
 ($R^2=0.13$ for PoPMuSiC and $R^2=0.14$  I-Mutant2.0(sequence+structure)) largely outperform those 
 who do not ($R^2\sim 0.02$ for MUpro and I-Mutant),  one gets only a modest further 
 improvement letting the mutated polypeptide relax via molecular simulations 
 ($R^2=0.17$ for molecular simulations, see Fig. \ref{fig:corr_bar}).

It is well known that residues buried in the protein core are important determinants of protein stability. Mutation affecting these sites tend to be highly destabilizing \mod{\cite{ponder1987tertiary,bustamante2000solvent,franzosa2009structural,abriata2015structural}}. Therefore, we test also to what extent solvent accessibility explains the experimental mutation effects. Upon defining
\begin{equation}
\Delta\phi_{\modminor{RSA}}(a_i\to b)=\alpha_i\ ,
\end{equation}
where $\alpha_i$ is the \modminor{relative solvent accessible surface area (RSA) of residue $a_i$ in position $i$. We use Michel Sanner's Molecular Surface (MSMS) algorithm \cite{sanner1996reduced} applied to the PDB structure 1M40 to estimate surface accessible surface areas (SAS), normalized by the maximum accessibilities given in \cite{tien2013maximum}. We find that $R^2=0.20$ of the variability of the experimental fitness is explainable via RSA. In general, 
we find that different accessibility estimates provide very similar results, including the absolute SAS, cf. the {\em Supplement}. Indeed, a simple binary classifier 
roughly distinguishing buried from exposed residues is almost as informative as RSA and SAS values (Fig.~S5).}
Note that the score $\Delta\phi_{\modminor{RSA}}$ does not depend on the target amino acid $b$, but only on the wild-type structure. Note also that this $R^2$-value, while been greater than those achieved through molecular simulations, is substantially smaller than all statistical sequence scores derived from homologs.

The failure of stability-based predictions of mutational effects may result from strong-effect mutations in or close to the active site, whose phenotypic effect is unrelated to protein stability. To assess this effect, we have repeated our analysis
including only 111 mutations falling into the extended active site, cf. the Supplementary Fig.~S6 for details. The $R^2$-values for both statistical models (IND and DCA) go up strongly ($R_{IND}^2=0.52, R_{DCA}^2=0.67$), while the structure-based predictors show little or no gain at all. This demonstrates, that evolutionary information accurately predicts the effects of mutation falling into the active site, and structural information does not.

Being grounded on complementary sources of information, predictions by evolution- and structure-based methods are not strongly correlated, as shown in Supplementary Fig.~S4. A linear combination of DCA with structural predictors, however, yields only little increase in correlation: the explained variance of experimental data gets to $0.60 \sim 0.61$ when performing a bivariate linear regression between DCA scores and either solvent accessibility or Polyphen-2 predictions, as displayed in Supplmentary Fig.~S7.

\subsection*{DCA landscape modeling spots stabilizing mutations and captures protein-specific substitution 
scores}

\begin{figure*}
\begin{center}
\includegraphics[width=.9\textwidth]{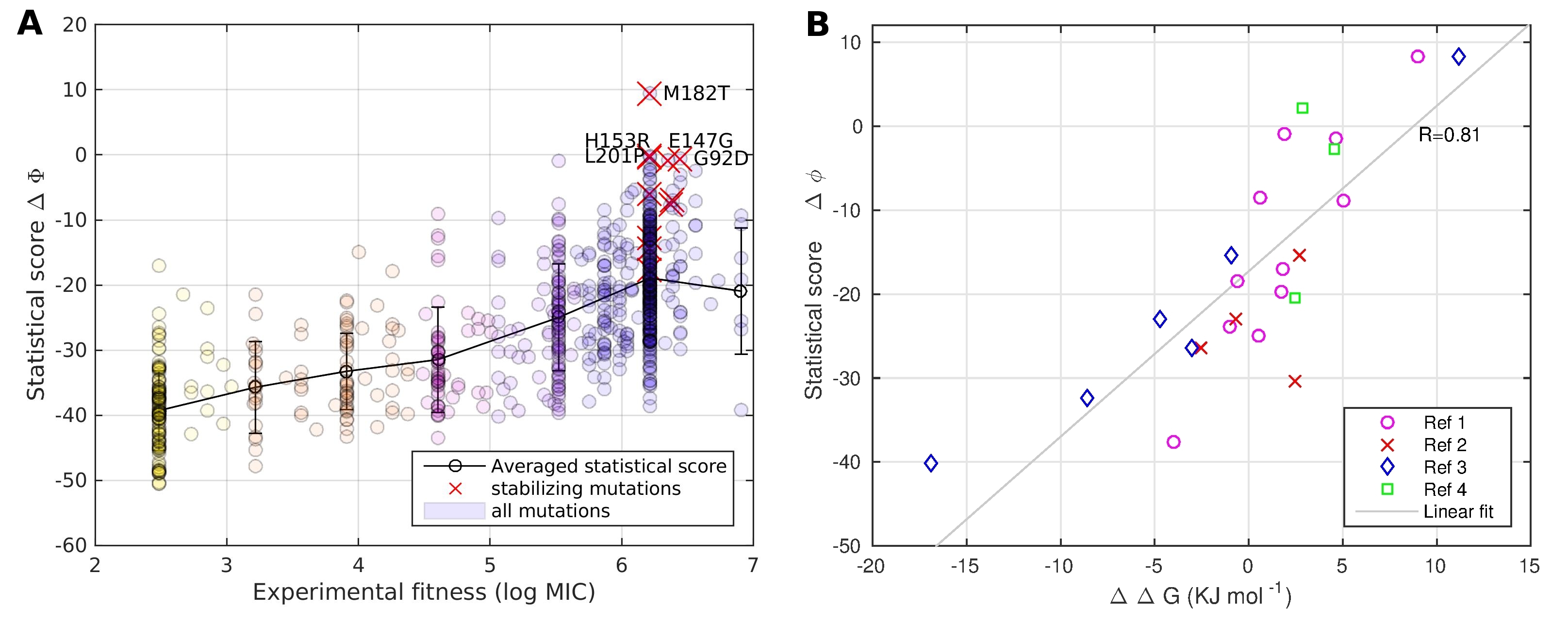}
\vspace*{0.05cm}
\caption{\label{Fig7bis} {\bf Statistical scores and thermodynamic stabilities} Panel A: Scatter plot of the log odd ratio $\Delta \phi_{DCA}$ vs the experimental fitness $\mu_{exp}$, stabilizing mutations mentioned in the text are highlighted in red. The highest scoring mutations are M182T, G92D, H153R, L201P and E147G, all reported as stabilizing. 
Panel B:  $\Delta \phi_{DCA}$ for a smaller set of single mutations is now plotted vs. the change in Gibbs Free Energy relative to wild type $\Delta\Delta G \equiv \Delta G_{mut}- \Delta G_{wt}$, as measured by four independent studies (Ref1, Ref2, Ref3 and Ref4 are \cite{kather2008increased},\cite{raquet1995stability},\cite{wang2002evolution} and \cite{deng2012deep} respectively).
}
\end{center}
\end{figure*}

The TEM-1 beta-lactamase has been the subject of intense studies with regard to 
protein structure, function, and evolution, and a number of structurally stabilizing
substitutions have been identified \cite{deng2012deep,kather2008increased,raquet1995stability,wang2002evolution}:  P62S, V80I, G92D, R120G, E147G, H153R, M182T 
(strongly stabilizing), L201P, I208M, A184V, A224V, I247V, T265M, R275L/Q, and N276D 
 (positions are indicated using standard Ambler numbering \cite{ambler1991standard}). Some of them were found to influence the 
resistance phenotype  \cite{salverda2010natural}. Notably, the five highest DCA scores 
$\Delta \phi_{DCA}$ out of all considered mutants belong to this set: M182T, H153R, E147G, L201P and G92D (with a large gap 
separating the likelihood of the strongly stabilizing M182T from the scores of the 
other four, cf.~Fig.~\ref{Fig7bis}).
More quantitatively, we found that the Gibbs Free Energy change relative to wild type $\Delta\Delta G$ of a different, small set of mutations (most of which not affecting Amoxicillin resistance) characterized by four independent studies \cite{kather2008increased,raquet1995stability,wang2002evolution,deng2012deep} are highly correlated with DCA scores ($R_{DCA}=0.81$) but less correlated when using independent model ($R_{IND}=0.62$).

\begin{figure*}
\begin{center}
\includegraphics[width=.9\textwidth]{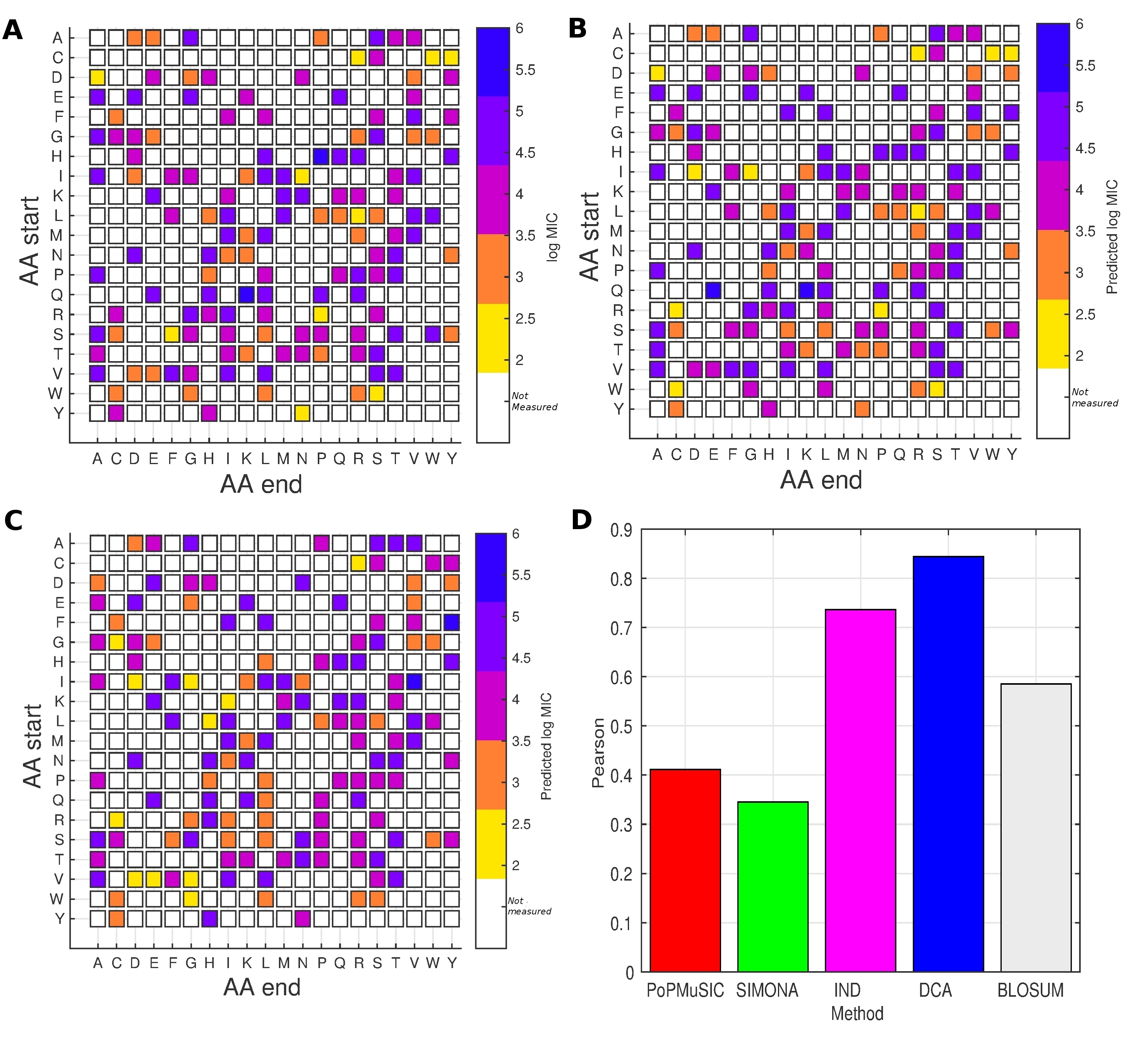}
\vspace*{0.05cm}
\caption{\label{Fig3} {\bf Protein-specific amino-acid substitution effects in TEM-1:} Amino acid substitution effects, averaged over experimental measurements (Panel A), DCA predictions (Panel B), and extracted from BLOSUM62 (Panel C). Blues squares correspond to nearly neutral mutations (log MIC $> 5.3$), while yellow squares correspond to highly deleterious mutations (log MIC $<2.6$). White squares are used for unobserved substitutions. The histogram in Panel D shows {\scriptsize $R^2$} between averaged computational and experimental amino-acid substitution effects.}
\end{center}
\end{figure*}

We further investigate whether the statistical analysis of homologous 
sequences is able to capture {\em protein-specific amino-acid substitution effects},
i.e.~if the effect of a specific amino-acid substitution (averaged over all sequence 
positions where this mutation appears) is better described by our statistical model
than it would be by Blosum matrices, which are estimated from many distinct
aligned protein sequences. To this aim, a matrix 
of average substitution scores is built from the set of experimental MIC values, 
cf.~Fig.~\ref{Fig3}. We also construct an analogous matrix for the 
DCA-predicted MIC values of the same set of mutations, and quantify
correlations between predicted and experimental average effects computing a 
Pearson correlation weighting each term with the square root of the number of 
measured mutations falling in the related class. We find a very large correlation
($R^2=0.72$) between average experimental and predicted substitution matrices.
This value has to be compared with the substantially lower correlation found
when comparing the mutational effects in TEM-1 with the Blosum62 
matrix ($R^2=0.34$), which provides amino-acid substitution scores averaged 
over many proteins. All other inference methods show substitution scores with
correlations to MIC, which are comparable to or lower than the correlations
between MIC and Blosum62.

\section*{Discussion\label{sec:Discussion} }

The central aim of this paper is the accurate computational inference of protein mutational landscapes to predict the phenotypic effect of mutations. This is exemplified in the case of the TEM-1 protein of {\em E. coli}, a beta-lactamase providing antibiotic drug resistance against beta-lactams, like penicillin, amoxicillin or ampicillin.

To reach this aim, we have extracted information about a protein and its potential mutants, which is hidden in the sequence variability of {\em diverged but functional} homologs of this protein. The central ingredient of our analysis is a careful modeling of residue coevolution by Direct-Coupling Analysis, i.e. the modeling includes pairwise epistasis between residues. This approach, initially developed in the context of structural biology in order to predict residue-residue contacts from sequences, has been used to define a score for each mutation, which was found to explain 55\% resp. 58\% of the phenotypic variability in the two corresponding experimental TEM-1 data sets \cite{jacquier2013capturing,firnberg2014comprehensive}. This value is substantially higher than what can be obtained by a more standard modeling approach based on sequence profiles (39\% of variability explained), which does not include epistasis, or on changes in structural stability. Furthermore, our coevolutionary approach clearly outperforms state-of-the-art approaches like SIFT and PolyPhen-2, which are based on non-epistatic models.

However, epistatic effects are not equally important for all residues, which may explain that some authors disagree on the contribution of the sequence context to mutational effect \cite{pollock2012amino,ashenberg2013mutational,zou2015convergent}.
The relevant context determining the effect of a mutation of a residue is not only given by its direct physical neighbors, but extends to a distance of about 20\AA. The informative context thus includes, on average, roughly half of all residues in the aligned TEM-1 sequence. \mod{This result agrees with the finding that interactions from second shell and beyond might be important for protein function \cite{drawz2009role}}.
 Having a look to the physico-chemical properties of the wild-type and the mutant amino-acids, we observe, e.g., that mutations substituting a hydrophobic residue with a hydrophilic one are almost equally well described by the DCA and by the independent model ($R_{DCA}^2-R_{IND}^2\simeq 5\%$), due to the structurally highly disruptive effect of a hydrophilic residue in a buried site, and thus the absence of hydrophilic residues in the corresponding column of the sequence alignment. On the contrary, the more moderate effect of replacing a small by a large amino acid depends strongly whether the context is able to accomodate this change or not, and thus the independent model performs much worse than the DCA model ($R_{DCA}^2-R_{IND}^2\simeq 26\%$). Concentrating on mutations from amino acids of given physico-chemical characteristics (hydrophobicity, charge, volume) toward a target amino acid of either different (e.g. hydrophobic to hydrophilic) or conserved characteristics (e.g. hydrophobic to hydrophobic) we find that the DCA predictions are stable, with $R^2$-values between 49 and 64\%, while the ones of the IND model vary much more strongly (25-55\%). In none of the considered cases, the independent model was able to outperform the coevolutionary one.

Our findings demonstrate that the {\em local mutational landscape} dictating the mutational effects in TEM-1 is closely related to the (co-)evolutionary pressures acting globally across the entire homologous protein family. This result is quite remarkable: Despite a low typical sequence identity of about 20\% between homologous beta-lactamases and TEM-1, their sequence statistics provides quantitative information about the effect of single-residue substitutions in TEM-1. We are thus able to infer landscapes and predict quantitatively mutational effects even in cases, where mutaganesis data are not sufficiently numerous, cf. \cite{Otwinowski03062014}. This complements recent findings, that patterns of polymorphism and covariation in patient derived (and thus highly similar) HIV sequences are informative about their replicative capacities \cite{shekhar2013spin,chakraborty2014hiv}, thanks to high mutation rates in the HIV virus. \mod{Further more, coevolutionary patterns in protein families were recently found to be closely related to protein energetics and folding landscapes \cite{lui2013network,morcos2014coevolutionary}}.


We expect that the modeling approach via DCA can be improved along several lines. First, prediction accuracy depends critically on the quality and size of the training multiple-sequence alignment. As we have shown, the prediction for gapped (and typically less well-aligned) positions is substantially worse than the one for ungapped (thus better alignable) ones ($R^2$-values raging from 30\% to 78\% from the most to the least gapped positions). We therefore excluded gapped sequences from the training alignment, but this procedure reduces the sequence number and thus the statistics for the ungapped positions.

Second, the current DCA approach is purely statistical and based on evolutionary information. It does not take into account any complementary knowledge about the protein under study. We have, however, observed that the integration of structural knowledge helps to increase the prediction accuracy. Fitting the model only for residues within about 20\AA~from the mutated residue, the $R^2$-value raises slightly by about 2\%. The effect of integrating the DCA-score and the solvent-accessible surface area is even larger, leading to a gain in $R^2$ of more than 6\%. A very similar increase (7\%) is obtained when combining DCA with PolyPhen-2, the latter being built upon a profile model and structural information. These increases are based on a simple linear regression scheme with threefold crossvalidation: It will be interesting to explore more sophisticated approaches, e.g. integrating prior structural knowledge via a Bayesian inference scheme directly into the statistical-inference procedure.

Even if the integration of complementary information may substantially improve our prediction accuracy, the most important contribution is, however, coming from the careful inclusion of epistatic effects into our modeling approach to mutational landscapes, \mod{as shown by a partial-correlation analysis in Fig.~S8}.

\mod{From a computational point of view, t}he approach is widely applicable beyond the specific case of TEM-1 and antibiotic drug resistance. \mod{To check this practically, we have analyzed further systems in the {\em Supplement}: a PDZ domain \cite{mclaughlin2012spatial}, a RNA recognition motif \cite{melamed2013deep} and the glucosidase enzyme \cite{romero2015dissecting}, cf. Supplementary Text S1 and Figs.~S9-S11.  DCA predictions systematically outperform independent-site models neglecting epistasis and all other tested methods. Only PolyPhen-2 reaches, in two cases out of four, comparable performance. Despite this encouraging finding, correlations between experiment and computation are numerically smaller than those observed for TEM-1. We expect this reduction to result from discrepancies between the measured phenotypes (e.g. protein stability, binding affinity) and those under evolutionary selection (fitness); MIC is without doubt a better proxy  for fitness than most molecular phenotypes. However, to systematically support this idea, large-scale experiments assessing the impact of mutations on multiple phenotypic traits in the same protein would be necessary. In summary, despite not representing a comprehensive survey, currently available data suggest a large potential for coevolutionary models in biomedical applications, via the {\em in silico} prediction of the role of mutations in rare diseases and cancer.}

\section*{Methods\label{sec:Methods}}

\subsection*{Data}

\subsubsection*{\label{sec_experimental}Mutational data}

The original dataset \cite{jacquier2013capturing} was used directly at the translated amino-acid level. It
contains $8621$ ($4094$ distinct) measurements of amoxicillin MIC. Among these  $8112$ do not include stop codons, $2440$ are repeated measures of the wild-type sequence,  3129 ($N_{multiple}=2051$ distinct) have all mutations inside the part of the sequence covered by the Pfam domain (i.e. subject to the presented statistical analysis). Finally, among the latter set, there are $N_{single}=742$ distinct single mutation. Each measurement $z_i$ falls in 9 discrete classes: $12.5, 25, 50, 125, 250, 500, 1000$, $2000, 4000$ (mg/L)  (no single point mutation has $z>1000$). For a given phenotype where amino acid $a_i$ in position $i$ is replaced with amino acid $b$ we have defined a unique experimental fitness $\mu_{exp}(a_i\to b)$ taking the logarithmic average on all measurements (whenever multiple measurements were available):

\begin{equation}
\mu_{exp}(a_i\to b)=\frac{1}{N(a_i\to b)}\sum_{i=1}^{N(a_i\to b)}log (z_i)
\end{equation}

where $N(a_i\to b)$ is the number of measurements of mutation $a_i\to b$.

\subsubsection*{Homologous sequences and preprocessing of the training set}

The genomic model was learned from a multiple sequence alignment (MSA) of sequences belonging to the Pfam Beta-lactamase2 family (PF13354) \cite{finn2013pfam}. We have used HMMer \cite{mistry2013challenges} to search  against the Uniprot protein sequence database (version updated to March 2015). The resulting MSA is $L=197$ sites long, and contains 5119 distinct sequences. After removing all sequences with more than 5 gaps, 2462 sequences are retained and used for the statistical analysis. They have an average sequence identity $\sim 20\%$ with the TEM-1 wild-type sequence.

\subsection*{Statistical sequence modeling}

\subsubsection*{Independent model -- sequence profile}

The basic assumption of the independent model Eq. (\ref{eq:IND}) is the additivity of the mutational effects of different positions in the amino-acid sequence. In terms of statistical sequence models, this corresponds to a
{\em sequence profile model}, which assigns to each sequence the factorized probability
\begin{equation}
P_{IND}(a_1,...,a_L)=\prod_{i=1}^L f_i(a_i) 
\label{eq:Pind}
\end{equation}
with $f_i(a)$ being the frequency of aminoacid $a$ in column $i$ of the MSA, see below for a precise definition of this frequency. The factorized form of this expression suggests to use log-probabilities as a computational predictor of the genotype-to-phenotype mapping,
\begin{equation}
\phi_{IND}(a_1,...,a_L) = \log P_{IND}(a_1,...,a_L) \ .
\end{equation}
This leads to an explicit expression of the phenotypic contribution of amino acid $a$ in site $i$: 
$\varphi_i(a) = \log f_i(a)$.

\subsubsection*{Epistatic model -- Direct-Coupling Analysis}

Following last paragraph's idea to identify the computational predictor of the genotype-to-phenotype mapping with the log-probability of a statistical model inferred from an MSA of TEM-1 homologs, the latter takes the form
\begin{equation}
P_{DCA}(a_1,...,a_L)=\frac 1Z \exp\{\phi_{DCA}(a_1,..,a_L) \} 
\end{equation}
where
\begin{equation}
\phi_{DCA}(a_1,...,a_L) = \sum_{i=1}^L \varphi_i(a_i)+\sum_{i,j>j=1}^L \varphi_{ij}(a_i,a_j)
\end{equation}
is given in Eq. (\ref{eq:Pdca}), and the so-called partition function 
$Z=\sum_{a_1,...,a_L} \exp\{\phi_{DCA}(a_1,..,a_L) \} $ is a normalization factor. The statistical model $P_{DCA}$ thus takes the form of a generalized Potts model or, equivalently, a pairwise Markov random field. The same model was introduced in the Direct-Coupling Analysis of residue coevolution \cite{weigt2009identification,morcos2011direct}. Inferring model parameters $\varphi$ from the MSA is a computationally hard task, we therefore follow the mean-field approximation introduced in \cite{morcos2011direct}. In this context, the epistatic couplings can be determined by inversion of the empirical covariance matrix $C_{ij}(a,b)$ for the co-occurrence of amino-acids $a$ and $b$ in positions $i$ and $j$ of the same protein sequence. Once the model parameters are determined, the context-dependent mutational effects can be estimated using Eq. \ref{eq:dcascore}.

{\color{black}
\subsubsection*{Details of statistical inference}

To take into account phylogenetic correlations and sampling biases in the training set,
each sequence $(a_1^m,...,a_L^m),\ m=1,...,M,$ of the MSA appears in the statistics 
with the following weight,
\begin{eqnarray}
w_m=\left( 1+ \sum_{m\neq m'}\theta\Big(d_{m,m'}-\vartheta L\Big) \right)^{-1} \ ,
\end{eqnarray}
with $d_{mm'}$ being the Manhattan distance (number of mismatches) between sequences $m$ and $m'$ and $\theta$ being the Heaviside step function whose value is zero for negative argument and one for positive argument. The reweighting threshold is set to $\vartheta=0.8$ as usually done in DCA \cite{morcos2011direct}).

Due to finite sampling, the statistics of the MSA has to be regularized introducing pseudocounts:
\begin{eqnarray} 
f^{\lambda}_{ij}(a,b)&=&\frac{\lambda}{q^2}+\frac{1-\lambda}{M_{eff}}\sum_{m=1}^M w_m \delta_{a_i^m,a}   \delta_{a_j^m,b}    \\
\quad f^\lambda_i(a)&=&\frac{\lambda}{q}+\frac{1-\lambda}{M_{eff}}\sum_{m=1}^M w_m \delta_{a_i^m,a}  
\end{eqnarray}
with $M_{eff}=\sum_{m=1}^M w_m$ and $\delta$ the Kronecker's delta whose value is one if the variables are equal, and zero otherwise.
We have included pseudocounts at two levels: First, for the inference of epistatic couplings we have used large pseudocounts ($\Lambda_2=0.5$), needed to correct for systematic biases introduced by the MF approximation  \cite{barton2014large},
\begin{eqnarray}
 \varphi_{ij}(a,b)&=&[C^{-1}]_{ij}(a,b)\\
 C_{ij}(a,b)&=&f^{\Lambda_{2}}_{ij}(a,b)-f^{\Lambda_{2}}_i(a) f^{\Lambda_{2}}_j(b)
\end{eqnarray}
for all amino acids $a$ and $b$. Following \cite{tanaka1998mean}, also diagonal terms
$\phi_{ii}(a,b)=[C^{-1}]_{ij}(a,b)$ are included. Couplings with gaps are set to zero, $\varphi_{ij}(a,-)=\varphi_{ij}(-,a)=0$, cf.~\cite{morcos2011direct}. 

Smaller pseudocounts of Bayesian size ($\Lambda_1=\frac{1}{M_{eff}}$) have been used in the regularization of single site frequencies  to infer the fields:
\begin{eqnarray}
\varphi_i(a)=\log\left(\frac{f_i^{\Lambda_{1}}(a)}{f_i^{\Lambda_{1}}(-)}\right)-\sum_{j,b} \varphi_{ij}(a,b)f_j^{\Lambda_1}(b)
\end{eqnarray}
The same small regularization $\Lambda_1=\frac{1}{M_{eff}}$ has been adopted in the independent-site model. 
}

\subsubsection*{Mapping scores to MIC values \label{sec:mu_comp}}

To compare computational predictions with experimental MIC values, we map computational scores 
$\Delta\phi(a_i \to b)$ into predicted MIC $\hat{\mu}(a_i \to b)$, by first sorting them and then associating 
to the $n_{th}$ highest score $\Delta\phi_{n_{th}}$ the $n_{th}$ highest experimental MIC value $\mu_{exp}(n_{th})$,
\begin{equation}
\hat{\mu}(\Delta\phi_{n_{th}})=\mu_{exp}(n_{th})\ .
\end{equation}
We subsequently compute linear correlations between the predicted MIC $\hat\mu$ and the experimental one $\mu_{exp}$, resulting in non-linear rank correlations between experimental fitnesses and raw computational scores $\Delta\phi$. 

This procedure has proved to be more robust than the standard Spearman rank correlations, because of the peculiar distribution of experimental data (bimodal with many repeated measures), and helpful to reduce the statistical weight of outliers (such as strongly destabilizing mutations in the distribution of $\Delta\Delta G$ predicted by molecular simulations). However, numerical values of Spearman correlations are in general not very different from those obtained by our procedure.

\subsection*{Structural stability predictions}

\subsubsection*{Bioinformatic predictors}

A list of predicted $\Delta\Delta G$ of {\it E. coli} TEM-1 protein point mutations for the web-based programs mentioned in the article have been downloaded from the SPROUTS database \cite{lonquety2009sprouts}.

\subsubsection*{Force-field based molecular simulations}

Computation of protein thermodynamic stability is computationally very demanding: A direct calculation of thermodynamic stability by molecular dynamics simulations implies the sampling of complete folding and unfolding events. This is presently infeasible for proteins of the size of TEM-1 (286 amino acids).
An alternative, less expensive approach to estimate mutational effects on pritein stability is to look for locally stable configuration performing small structural relaxations from a reference structure, with the wild type amino acid replaced by the mutant amino acid. Assuming that the protein can be described by a two-state system (folded vs. unfolded), and that both the entropy of the folded and the free energy of the unfolded are not sensibly affected by the mutation, we can approximate
\begin{equation}
\Delta \Delta G  \equiv \Delta G_{wt} - \Delta G_{mut} \simeq E_{wt}^{fold}-E_{mut}^{fold} \equiv \Delta E\ .
\end{equation}
Moreover, as thermodynamic stability is an equilibrium property, one can replace expensive molecular-dynamics simulations with more efficient Monte-Carlo sampling.

Molecular simulations were performed using SIMONA \cite{strunk2012simona}, a Monte-Carlo based simulation software for efficient molecular simulations which have proved useful to obtain reproducible folding in a series of test cases \cite{schug2003reproducible,verma2006basin}.
As reference structure for molecular relaxations we have taken a highly resolved (0.8\AA) structure (PDB: 1M40 \cite{minasov2002ultrahigh}). Further details of the simulations are reported in next section.

{\color{black}
\subsubsection*{Details and calibration of the molecular simulations}

To estimate the thermodynamic stability of TEM-1 mutants we have executed the following steps:
\begin{enumerate}
\item  Starting from a sufficiently close reference state (in our case the SIMONA-relaxed structure of the wild type molecule), the wild-type amino acid is replaced by the mutant one.
\item Monte-Carlo simulations are performed under SIMONA, to locally minimize the energy function.
\item The resulting energy change $\Delta E=E_{mut}-E_{wt}$ is determined.
\end{enumerate}
In the simulation, we have included the complete forcefield PFF03v4-all parallel OpenMP (scale 1.0), which makes use the  amber99sb-star-ildn dihedral potential with an implicit solvent model. It contains the following contributions:

\begin{eqnarray}
V(\{\vec{r_i}\})&=&\sum_{ij}V_{ij}\left[\left(\frac{R_{ij}}{r_{ij}}\right)^{12}-2\left(\frac{R_{ij}}{r_{ij}}\right)^6\right]+ \nonumber \\
&+&\sum_{ij}\frac{q_iq_j}{\epsilon_{g(i)g(j)} r_{ij}}+\sum_i \sigma_i A_i+ \sum_{hBonds}V_{hb}.\nonumber
\\
\end{eqnarray}
where $r_{ij}$ represents the distance between atoms $i$ and $j$, and $g(i)$ the type of amino-acid $i$, $V_{ij}$ and $R_{ij}$ are Lennard-Jones parameters, $q_i$ and $\epsilon_{g(i)g(j)}$ are the partial charges and group-specific dielectric constants for non trivial electrostatic interactions, $\sigma_i$ and $A_i$ are the free energy per unit area and the area of atom $i$ in contact with fictitious solvent respectively, and finally $V_{hb}$ is a short range interaction term for backbone-backbone hydrogen bonding \cite{schug2003reproducible}.
}

\section{Supplementary Material}
Supplementary Tables S1-S3, Figures S1-S13, Texts S1 and a Matlab implementation of DCA modeling and sequence scoring are available  at Molecular Biology and Evolution online (http://www.mbe.oxfordjournals.org/).

\section*{Acknowledgments}

We are grateful to Jacques Chomilier for help with the SPROUTS database. MW was partly funded by the Agence Nationale de la Recherche project COEVSTAT (ANR-13-BS04-0012-01). This work undertaken partially in the framework of CALSIMLAB is supported by the public grant ANR-11-LABX-0037-01 overseen by the French National Research Agency (ANR) as part of the "Investissements d'Avenir" program (ANR-11-IDEX-0004-02).

\bibliographystyle{unsrtnat}
\bibliography{bib_dca.bib}

\end{document}